\documentclass[final]{IEEEtran}
\usepackage{algorithm}
\usepackage{color}
\usepackage{algpseudocode}
\usepackage{graphics}
\usepackage{epsfig}
\usepackage{cite}
\usepackage{threeparttable}
\usepackage{graphicx}
\usepackage{picinpar}
\usepackage{amssymb}
\usepackage[cmex10]{amsmath}
\usepackage{url}

\usepackage{bm}
\usepackage{setspace}

\input epsf

\newcommand{\tabincell}[2]{\begin{tabular}{@{}#1@{}}#2\end{tabular}}
\begin{document}

\title{\huge{Near Space Communications (NS-COM): A New Regime in Space-Air-Ground Integrated Network (SAGIN)} }

\author{Zhenyu Xiao, \IEEEmembership{Senior Member,~IEEE}, Tianqi Mao, \IEEEmembership{Member,~IEEE}, Zhu Han, \IEEEmembership{Fellow,~IEEE}, and Xiang-Gen Xia, \IEEEmembership{Fellow,~IEEE}
\thanks{Zhenyu Xiao and Tianqi Mao are with Beihang University; Zhu Han is with University of Houston; Xiang-Gen Xia is with University of Delaware; The corresponding author is Tianqi Mao.} %
\vspace{-3mm}} %

\maketitle
\begin{abstract}
Precipitated by the technological innovations of the near-space platform stations (NSPS), the near space communication (NS-COM) network has emerged as an indispensable part of the next-generation space-air-ground integrated network (SAGIN) that facilitates ubiquitous coverage and broadband data transfer. This paper aims to provide a comprehensive overview of NS-COM. Firstly, we investigate the differences between NS-COM and the existing terrestrial cellular networks as well as satellite-based and unmanned-aerial-vehicle (UAV)-based communication networks, which is followed by a review of the NS-COM development. Then, we explore the unique characteristics of NS-COM regarding the platforms and the propagation environment of the near space. The main issues of NS-COM are identified, resulted from the extremely long transmission distance, limitations of the communication payloads on NSPS and complex atmospheric constitution of the near space. Then various application scenarios of NS-COM are discussed, where the special technical requirements are also revealed, from the physical-layer aspect like transceiver design to the upper-layer aspect like computational offloading and NSPS placement. Furthermore, we investigate the co-existence of NS-COM and ground networks by treating each other as interferers or collaborators. Finally, we list several potential technologies for NS-COM from the perspective of spectrum usage, and highlight their technical challenges for future research.
\end{abstract}

\section{Introduction}
\begin{figure*}[t!]
	\begin{center}
		\includegraphics[width=0.8\linewidth, keepaspectratio]{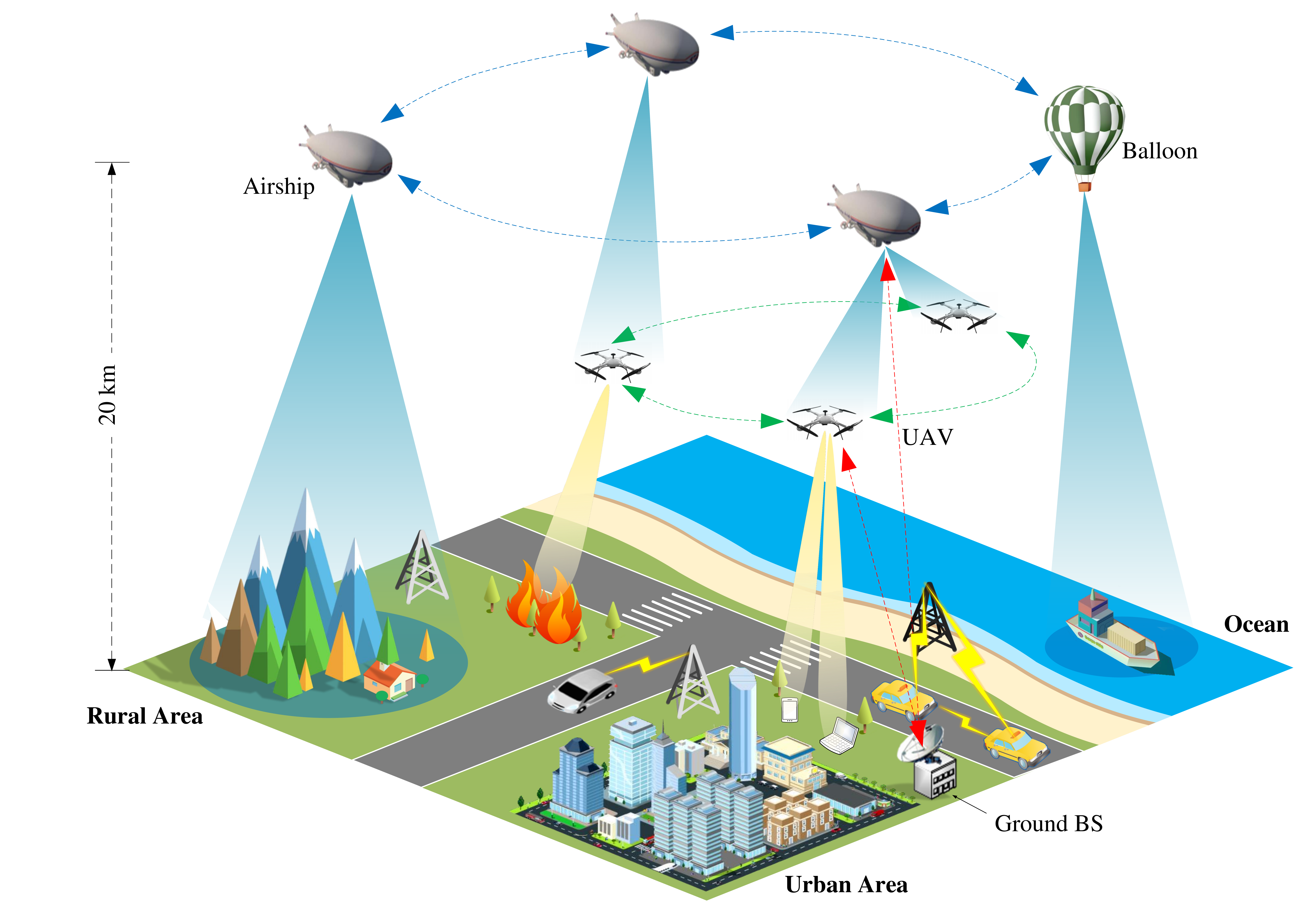}
	\end{center}
	\caption{Schematic diagram of the NS-COM networks.}
	\label{fig1}
\end{figure*}

The near space is referred to as the region ranging from 20 to 100 km above the ground, which contains most of the stratosphere, the mesosphere and part of the ionosphere. Thanks to its environmental peculiarities, including high radiation, low temperature, low moisture, low atmosphere pressure, etc., the near space has become a natural laboratory for researches on geobiology, astrobiology, physics, and meteorology. Besides, its height over troposphere also enables weather forecast applications and military reconnaissance with desirable stealthiness. Despite its significant strategic value, the exploitation of near space has stagnated for a long time, since the airplanes cannot adapt to such thin atmosphere, whilst low-earth-orbit (LEO) satellites are also inapplicable due to the gravity \cite{Zhang_conf_07}. This motivates rapid development of the near-space platform station (NSPS) technology in the past 20 years. State-of-the-art NSPS can be classified into aerostatic platforms relying on buoyancy of the lifting gas, and aerodynamic aircraft flying with aerodynamic lift \cite{Abbas_procIEEE_11}. With the aid of these NSPS constellations, the near-space communication (NS-COM) network can be established by building broadband data links among the NSPSs, and between the NSPS and the aerial or terrestrial user terminals, as illustrated in Fig. \ref{fig1}. Note that the aerostatic NSPSs are capable of keeping nearly geostationary in the air for several weeks or months with heavier payloads, which are more suitable for NS-COM than the aerodynamic counterparts. These aerostatic NSPSs consist of airships and balloons. Whilst the former enables flight control functions with the electric propeller and motor, the latter is more cost-efficient, making it preferable for massive deployment.

\begin{table*}[t!]
	\caption{Characteristic Comparison of the NS-COM Network with the Terrestrial Wireless, Satellite and UAV Counterparts {\cite{Alam_cm_21,Liu_survey_18}.}}	
	\centering
	\resizebox{0.8\linewidth}{!}{
		\begin{tabular}{|c|c|c|c|c|}
			\hline
			Issue & Terrestrial & Satellite & UAV & NS-COM \\\hline
			Operating duration & Permanent & Permanent & Short period & Long period \\\hline
			Throughput & High & Low & High  & High\\\hline
			Coverage & Hotspot area & Global area  & Local area & Wide area \\\hline
			Propogation Delay  & Low & High & {Low} & Moderate\\\hline
			Expense	& \tabincell{c}{Generally low,\\high in rural area} & High & Low & Low\\\hline
			Purpose & Commercial use & Commercial/Special use & Special use & Special use\\\hline	
	\end{tabular}}
	\label{t1}
\end{table*}

There exists much difference between the NS-COM network and classical space/air/ground networks regarding the communication requirement, as summarized in Table \ref{t1}. It is seen that NS-COM can incorporate the high throughput and low-cost deployment of airborne networks as well as the wide coverage of satellite networks. Despite the merits, the massive deployment of NSPS was limited by harsh requirements on the volume and material of the envelope \cite{Chan_cm_16}. Hence, the majority of research interests in the past were concentrated on the link/access techniques for NS-COM constituted by a limited number of (or even one) NSPS nodes\footnote{In most of the existing literature, the NSPS nodes are also referred to as high altitude platform stations (HAPSs) at an altitude of $20$ to $50$ km.}. For instance, {in the 1990s, NSPS was originally employed for fixed services, which could support high-speed point-to-point (P2P) connectivity to a fixed gateway or fixed customer premises equipment (CPE), aiming at building feeder links to the core network.} Besides, NSPS was utilized as the macro base station, not only to provide temporary coverage for unpredictable emergencies, such as network congestion and natural disaster, but also to realize longstanding data connectivity with both terrestrial and aerial user terminals with ubiquitous coverage \cite{Alam_cm_21}. 

In recent years, the growing thirst for full-utilization of the near-space resources has motivated rapid advancements of the NSPS technology, substantially enhancing its capabilities of both massive deployment and payload carrying. The breakthrough can be represented by the X-station \cite{StationX_2022}, HAPSMobile and Loon projects \cite{Loon_Google_2022}. Explicitly, commercial airships at an altitude of $21$ km with super-strength and super-light materials have been realized by StratXX, which were capable of providing various communication, positioning and navigation services with full coverage of $10^6$ $\text{km}^2$. {Besides, the SoftBank Corporation and its subsidiary HAPSMobile managed to develop a high-altitude autonomous tethered balloon base station system to enable reliable and high-quality connectivity for future Internet services. A footprint fixation technology by adjusting the cylindrical phased-array antennas of the balloon, was trialed to mitigate the impacts of possible motion and rotation of NSPS, so that the stability of Internet coverage could be guaranteed.} Besides, the Google-X spent more than nine years managing to develop mature balloon products that achieve 300 days' continuous fly in the stratosphere, and deployed thousands of them equipped with advanced solar panels and communication payloads in order to constitute a mesh network in the sky, aiming at providing reliable and affordable Internet services for billions of people. {Thanks to these advancements, the NS-COM network has become a promising complement to ground-based wireless communications, which can be incorporated with the classical terrestrial cellular network to achieve both high throughput and desirable coverage simultaneously. Despite the attractive merits, the integrated near-space-ground network also brings a plethora of new technical issues, such as interference mitigation, mobility management, handover management, traffic offloading strategies and routing algorithm design, which demand extensive investigation \cite{Liu_survey_18}.}

To provide the researchers with a more comprehensive understanding of NS-COM and inspirations of further advances, in this article, we firstly investigate the hardware and environmental characteristics of NS-COM. Then we discuss several representative application scenarios, including the near-space access network, the near-space mesh network as well as their co-existence with the ground communication systems, and the corresponding technical requirements are also specified. Finally, the candidate techniques for NS-COM as well as some open issues therein are highlighted, focusing on various frequency bands from sub-6 GHz to the optical scale. {Compared with existing relevant survey papers, this article concentrates more on the near-space mesh network that ensures ubiquitous coverage for 6G and beyond communication networks. We also focus on the benefits and challenges of several potential cutting-edge technologies, including the metasurface-based antenna array and terahertz (THz) wireless communications, to support the implementation of NS-COM for 6G and beyond communications in the future.}

\section{Peculiarities of NS-COM}\label{S2}
In comparison with the other existing space, air, or ground-based communication systems, the peculiarities of NS-COM can be generally classified into two aspects regarding the characteristics of NSPS as well as the unique propagation environment of the near space.



{\bf Characteristics of the Near-Space Platforms}: {Different from the existing terrestrial, satellite, or UAV networks, NSPS is expected to support broadband data transmission of the distance over $100$ km, yielding tough requirements on communication bandwidth and the antenna gain. On one hand, conventional sub-6 GHz spectrum has great difficulties in supporting bandwidth-consuming services due to its scarcity of available frequency resources. To address these issues, the World Radiocommunication Conference in 2019 (WRC-19) has allocated the millimeter-wave (mmWave) spectrum to worldwide NS-COM including the 31-31.3 GHz, 38-39.5 GHz, 47.2-47.5 GHz and 47.9-48.2 GHz bands, to provide sufficient communication bandwidth. On the other hand, due to its superior payload carrying capability as much as $1,000$ kg, NSPS presents stronger computational power than the other aerial platforms, which is qualified for computational offloading from aerial user terminals, e.g., UAVs. However, empirically, only tens of kg are allocated for communication purposes on NSPS, making it impractical to employ high-gain parabolic antennas as the terrestrial base stations. As a promising alternative, large-scale phased antenna arrays at mmWave frequencies can be employed on NSPS to enable highly directional inter-NSPS or NSPS-to-ground data transmission, which meet the limited budget for communication payloads, due to compact physical size of the mmWave-band antenna elements. Nevertheless, the resultant narrow beamwidth poses more stringent requirement accuracy of beam alignment, leading to severe antenna misalignment fading under mobility of NSPS caused by winds. Besides, the broadband data transmission at mmWave frequencies also adds to the hardware impairment of the frequency multiplier and the power amplifier. Such impairment can be even more pronounced for NSPS-mounted transceivers, since cost-efficient front-end payloads with light weight and compact size is preferred for NSPS under the constraints of communication payload.}

	\begin{figure}[t!]
	\begin{center}
		\includegraphics[width=1\linewidth, keepaspectratio]{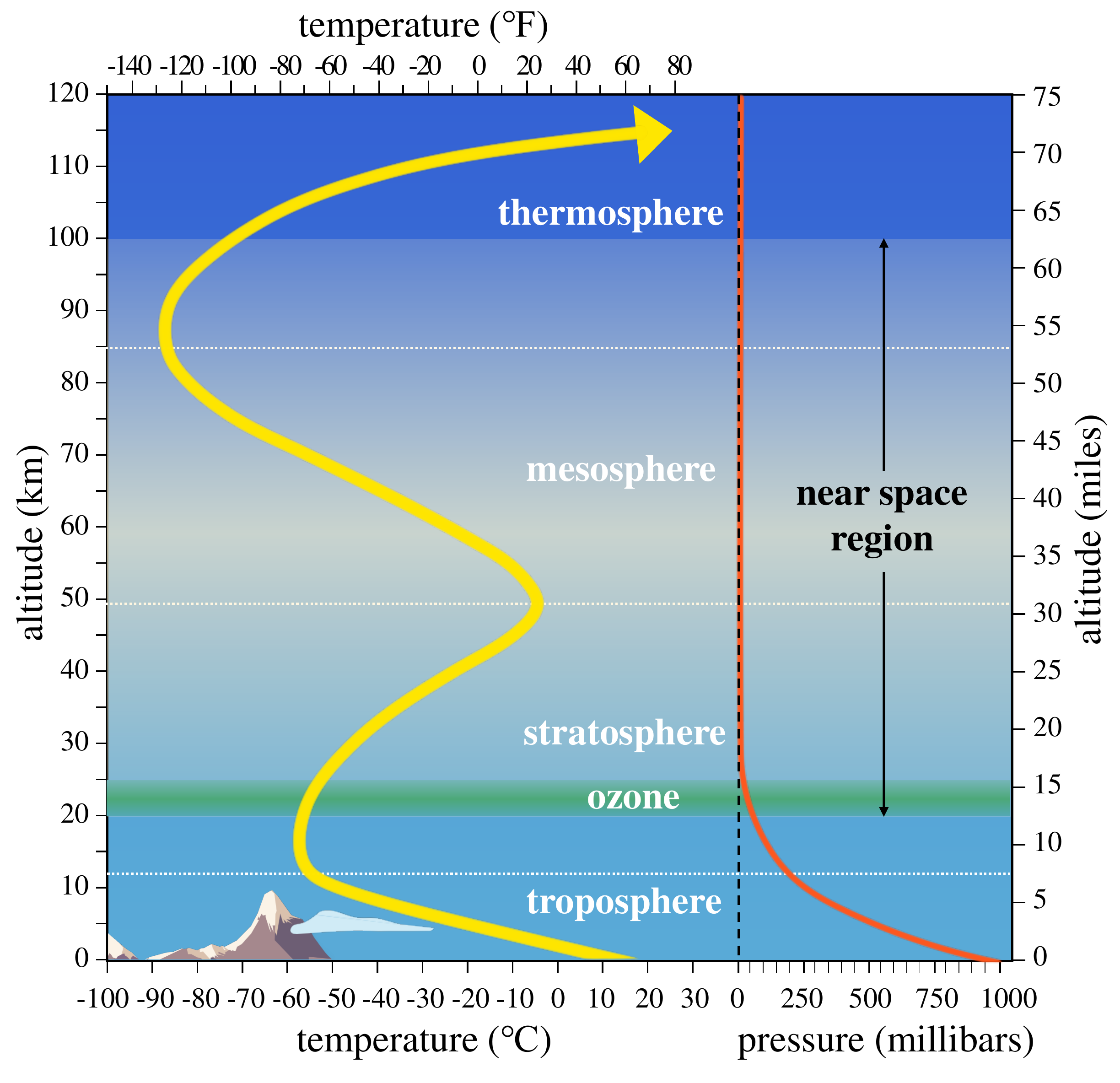}
	\end{center}
	\caption{Different layers of the Earth atmosphere, where the near-space region is specified, and the curves of atmospheric temperature and pressure with respect to the altitude are presented \cite{britannica_2022}.}
	\label{fig2}
	\vspace{-3mm}
\end{figure} 

{\bf Characteristics of the Propagation Environment}: Aside from the peculiarities of NSPS, the propagation environment of the near space is also completely different from those at other altitudes. As shown in Fig. \ref{fig2}, the near-space region contains the majority of the stratosphere, the mesosphere and part of the thermosphere, where the constituent of the air, temperature, and the atmospheric pressure are much different \cite{Ozdemir_conf_13}. Explicitly, aside from the trend to become thinner with higher flying altitude, the constituent of the atmosphere varies across different layers of the near space. In particular, the ozone gas in the atmosphere is mainly concentrated in the stratosphere, which presents a superior capability of absorption of the ultra-violet radiations. This further leads to a {non-monotonic} changing pattern of the temperature with respect to the altitude. For instance, the temperature in the near space is generally much lower than zero centigrade, whilst can approach zero at the top of stratosphere thanks to the strong radiation yet absorbed by the ozone layer. {Furthermore, the gradual changes of temperature and density of the atmosphere also cause non-identical atmospheric pressure. These complex factors of the near-space propagation environment make existing channel models of the airborne and satellite networks no longer applicable. In addition, refraction cannot be overlooked due to the non-homogenity of the transmission medium, which can bend the beam direction together with the scattering effects by aerosols, especially for the near-space-ground data links. This adds great difficulties to NS-COM beam alignment, causing frequent disconnections that hinder both the communication and computational offloading services.}



\section{Typical Application Scenarios and Requirements}\label{S3}
This section lists different application scenarios of NS-COM based on its unique characteristics, and the corresponding technical requirements are then discussed.

{\bf Near-Space-to-Ground P2P Link}: The unique characteristics including high radiation, extremely low temperature and atmospheric pressure, as well as sparse density of water vapor, {have made} the near space a desirable laboratory for biology, physics and meteorology. In addition, the near-space platforms are naturally desirable for observation of the typhoon phenomenon, since the majority impacts of typhoon are below the top of troposphere. These scientific exploration tasks would generate huge amount of observation data, which necessitates broadband P2P data transmission between NSPS and the ground control station. In order to guarantee both high data rate and reliability of the near-space-to-ground communication link over $100$ km, the mmWave communication system is more suitable than the sub-6 GHz counterparts with broadcasting nature, due to its attractive advantages including large available bandwidth and capability of maintaining highly directional beams via massive antenna array. {Besides, multi-hop relaying transmission is also a promising option to further enlarge the communication distance. This can be realized by the utilization of UAV relays, which has been demonstrated to support beyond-the-horizon transmission with desirable communication quality by incorporating various ``icing-on-the-cake'' technologies, e.g., the reconfigurable intelligent surface (RIS) \cite{Pang_wc_21}. 

\begin{figure}[t!]
	\begin{center}
		\includegraphics[width=0.8\linewidth, keepaspectratio]{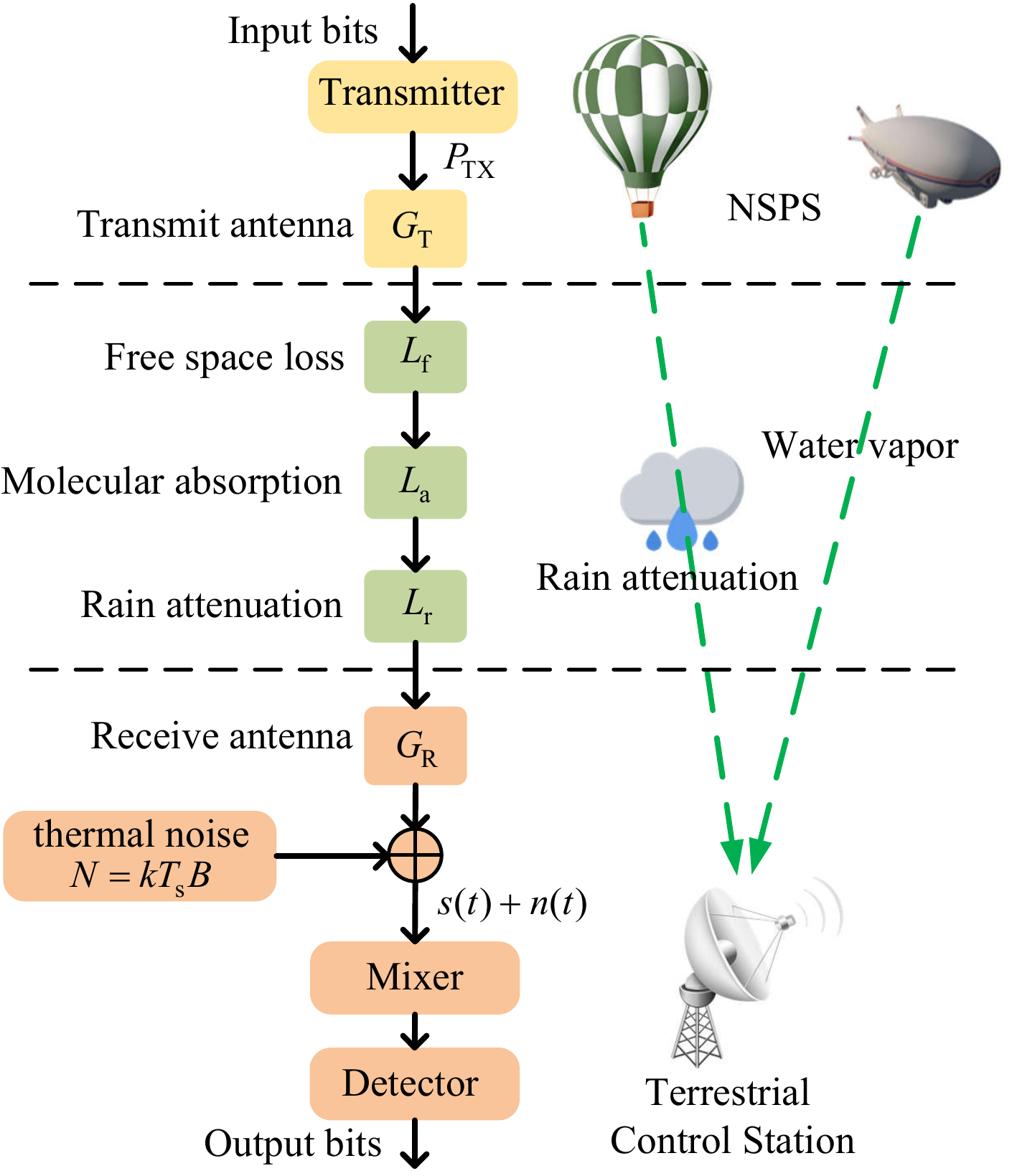}
	\end{center}
	\caption{Schematic diagram of the near-space-to-ground P2P link budget.}
	\label{fig3}
	\vspace{-3mm}
\end{figure} 	

On the other hand, the resultant issues of the aforementioned technologies cannot be overlooked. Explicitly, unlike the space environment, the wind effect in the near space approaches 30 knots at the altitude of 24 km and the latitude from 60 to 90 degrees}, leading to movement of NSPS relative to the Earth, as well as the orientation change of the antenna array. {These phenomena add great difficulty to the transceiver beam alignment, especially for mmWave directional data links, not to mention the aforementioned relaying schemes with high-mobility UAV platforms.} Therefore, on one hand, {the flight control subsystem should be carefully designed to compensate for the movement and orientation change, and maintain reliable data links of telemetry, tracking and commanding signals to further assist beam alignment between NSPS and the ground control station.} On the other hand, robust beamforming and adaptive beam tracking schemes are necessary, which should fully utilize the prior knowledge of the NSPS mobility and environmental information from the sensors, to achieve higher efficiency of the beam management.

For practical establishment of the near-space-to-ground data link, it is essential to perform the link budget analysis in order to determine the critical system parameters, such as the modulation order, the transmit power and the size of antenna array, which is especially necessary for mmWave communication systems due to strong rain attenuation and molecular absorption effects by water vapor. {Explicitly, the received signal-to-noise ratio (SNR) of the near-space-ground P2P link can be calculated according to the link budget shown in Fig. \ref{fig3}. In this schematic diagram, $P_{\text{TX}}$ denotes the average transmit power, and $G_\text{T}$ and $G_\text{R}$ denote the transmit and receive antenna gains, respectively. Besides, $L_\text{f}$, $L_\text{a}$ and $L_\text{r}$ represent the free-space path loss, molecular absorption, and rain attenuation. Additionally, $k$, $B$ and $T_\text{s}$ stand for the Boltzmann's constant, the communication bandwidth and the noise temperature of the receiver, respectively.}

\begin{figure}[t!]
	\begin{center}
		\includegraphics[width=1\linewidth, keepaspectratio]{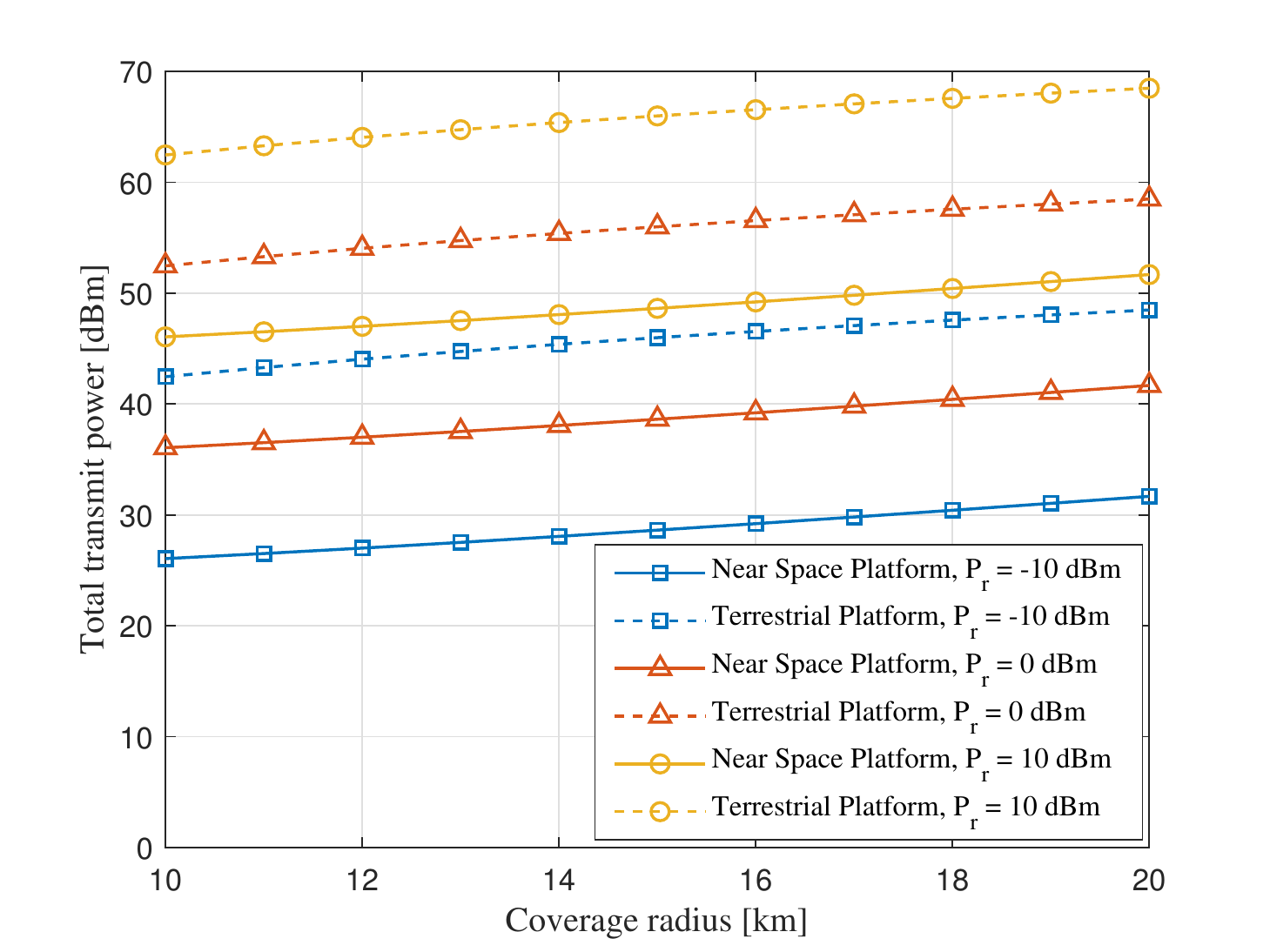}
	\end{center}
	\caption{Comparison of power consumptions between NS-COM and the classical terrestrial cellular networks with respect to the coverage radius, under different values of received power at the cell edge user.}
	\label{fig0}
	\vspace{-3mm}
\end{figure} 

\begin{figure*}[t!]
	\begin{center}
		\includegraphics[width=1\linewidth, keepaspectratio]{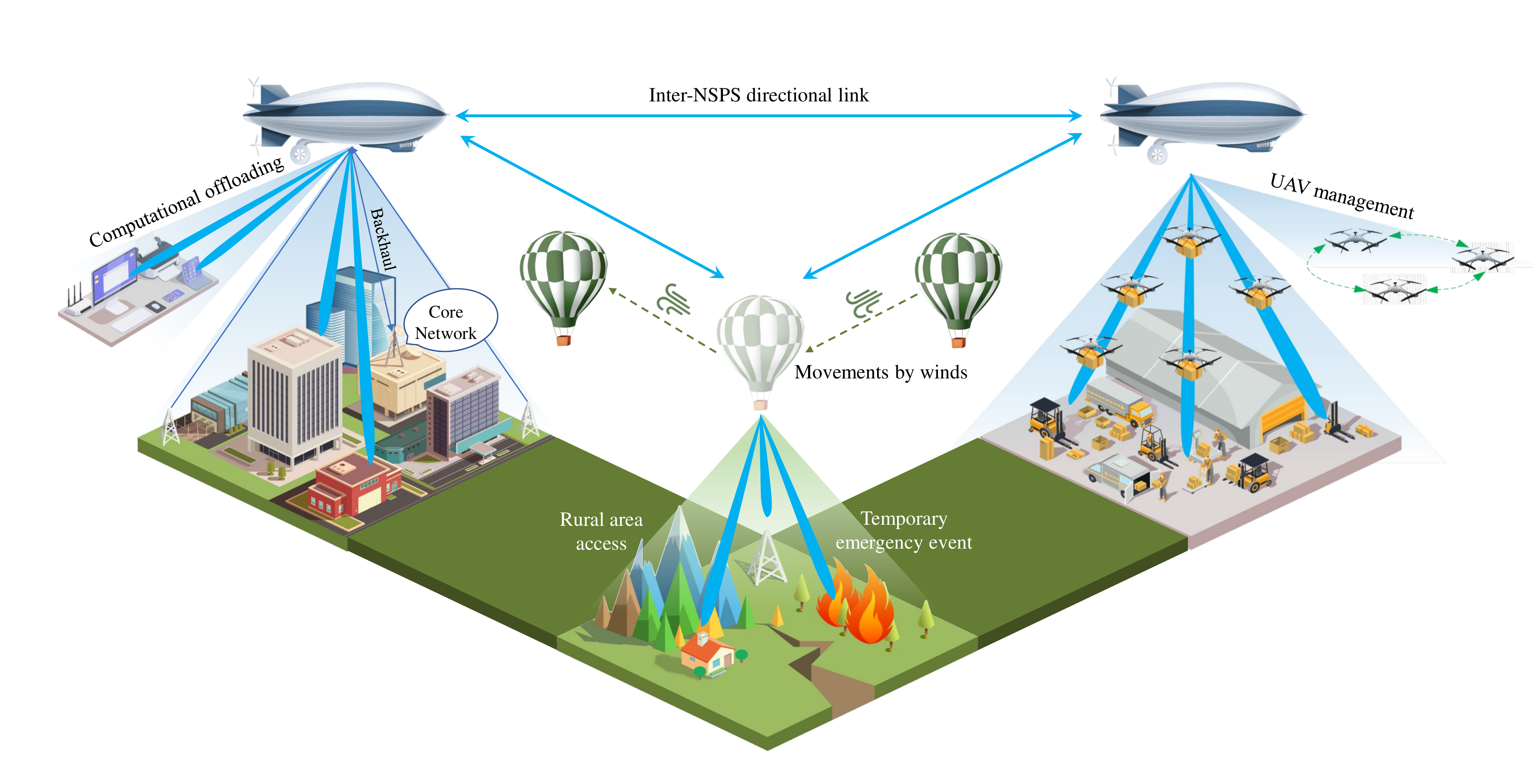}
	\end{center}
	\caption{Illustrations of the near-space access network and mesh network, where different applications of the near-space access network are presented, including Internet access for distant and disaster regions, dual-hop wireless backhauling, offloading of data traffic or computational tasks and management for UAV fleet and cargo drones. }
	\label{fig4}
\end{figure*} 

{\bf Near-Space Access Network}: {The NSPS has presented its superiority in different aspects, such as availability of line-of-sight (LoS) transmission, large footprint with favorable capacity, and low mobility. Thanks to these significant merits,} the near-space access network can be established based on the NSPS macro base station, which acts as a powerful complement to the fifth-generation (5G) and beyond wireless communications. {For instance, Fig. \ref{fig0} compares the power consumptions of near-space access networks and classical terrestrial cellular networks at different values of coverage radius, where equal power threshold of the received signals at ground cell-edge users is assumed, denoted as $P_\text{r}$. In simulations, the 2-GHz frequency band is employed for both NSPS at the altitude of $20$ km and terrestrial base stations. It can be observed that, the near-space access network is more energy efficient than its terrestrial counterpart under different requirements on the coverage and user experience. This is because NS-COM can provide reliable LoS links, whilst the terrestrial data transmission are more fragile to shadowing. Aside from high energy efficiency,} the near-space access network is capable of addressing numerous challenges of the existing terrestrial cellular network, as shown in Fig. \ref{fig4}. To elaborate more, NSPS can be easily deployed with relatively low expenses in rural areas as well as the disaster regions of tsunamis, hurricanes and earthquakes, enabling reliable Internet access. Besides, wireless backhauling has become a promising solution to small base stations in cellular networks due to its low deployment cost, which however, suffers from unavailability of the LoS paths especially in urban canyons. This can be addressed by the application of NSPS macro base stations to build dual-hop LoS links between small base stations and the core network. Moreover, the NSPS macro base stations usually possess powerful computational capability thanks to its sufficient volume for computing payloads, which can be utilized to offload the data processing tasks from the multi-media terminals, enabling a plethora of computation-consuming applications like virtual/augment realities. This desirable processing power can also be exploited for management and data offloading of the UAV fleet, whose computational ability is limited.

To guarantee the performance of the near-space access network under these application scenarios, several technical issues are required to be addressed. Firstly, the NSPS macro base station aims at providing the Internet services to both the airborne and ground users within an ultra-wide coverage of several tens of km in radius. The resultant sparsity of user distribution in the macro base station necessitates creating disjoint transmit beams for various user terminals located far away from each other. {Under such conditions, the beam widths and gains need to be optimized for overall capacity enhancement, based on different advanced tools like dynamic programming and artificial-intelligence (AI) methods. Besides, spectrum sharing can be invoked between different users to fully exploit the sparsity of their spatial distribution. The agile beam management operations can be guaranteed by massive multiple-input multiple-output (MIMO) technology.} However, in classical architecture based on phased-array antennas, a large number of combiners, power dividers, phase shifters as well the microstrip lines significantly increase the volume and weigh of the transceiver hardware, causing difficulty of payload arrangement and additional energy consumption for flight control. To tackle this issue, the RIS technology, also referred to as programmable meta-surface, can be introduced, which relies on a cost-efficient planar surface of enormous programmable passive reflecting elements \cite{Tang_wc_20}. Unlike the classical concept for communication relay applications, the meta-surface could replace the classical phased-array antenna at the transmitter, which is capable of performing analog precoding without feeding network. Consequently, the pressure on communication payloads of NSPS can be alleviated. 

Furthermore, elaborate design of the waveform and multiple access techniques for the near-space access network is necessary to provide long-distance broadband data access with satisfactory quality of service (QoS) levels for numerous users in the super macro cell. Major concerns include attaining high signal quality and spectral and energy efficiencies. Orthogonal frequency division multiplexing (OFDM)/Orthogonal frequency division multiple access (OFDMA) is mainly adopted in 4G, 5G and beyond communication networks thanks to simple equalization and robustness to time-dispersive channels. However, the time-domain OFDM signals inherently suffer from the high peak-to-average power ratio (PAPR) issue. This leads to clipping distortions induced by the power amplifier, causing degradation of both energy efficiency and error performance. Besides, the growing hardware complexity of OFDM/OFDMA with the increase of communication frequency and bandwidth, adds much burden to the near-space platform, especially when reaching sub-mmWave frequencies over 100 GHz. Since the frequency-selective channel fading is marginal for near-space access network due to the dominance of LoS near-space-to-ground paths, the single-carrier waveform might be preferable to OFDM for low PAPR and simple implementation, and the spatial division multiple access (SDMA) may fit quite well with the near-space access network due to its sparsity of the user distribution.


{\bf Near-Space Mesh Network}: {As presented in Fig. \ref{fig4}, the NS-COM coverage can be extended with a near-space mesh network established by interconnections among multiple near-space airships and balloons.} The near-space mesh network necessitates broadband long-distance transmission over 100 km between NSPSs. For the near-space-to-ground communications, high-gain parabolic antennas could be employed at the ground base station for path-loss compensation. However, it is inapplicable for near-space platforms due to limited payload budget for communication purpose. Instead, the inter-NSPS links for near-space mesh network have to employ highly directional beams at the transceiver, which is difficult to achieve with the classical sub-6 GHz spectrum. Fortunately, due to sparsity of water vapor above the stratosphere, the sub-mmWave, terahertz (THz) and even optical frequencies could be exploited to generate extremely narrow beams without much attenuation by the atmosphere, which will be discussed in Section \ref{S4}. 


For near-space mesh networking based on nearly geostationary airships, the multi-NSPS placement issue needs extensive investigation to enhance the overall performance of NS-COM, consisting of communication capacity, coverage and energy efficiency, etc. For instance, the NSPS topology can be improved to reach a desirable trade-off between the conflicting capacity and coverage of the near-mesh network \cite{Gong_tec_17}. By considering these two performance metrics as dual objectives, a multi-objective NSPS-placement optimization problem can be formulated, where various critical parameters of NS-COM should be taken into account, including the path loss, the routing efficiency in the mesh network, the safety issue, and the ground user distribution. Furthermore, the placement and power allocation strategies of the near-space mesh network should be adaptive to time-varying information of the terrestrial users, e.g., the densities of the user terminals or data traffic, in order to maximize the system energy efficiency. 

{On the other hand, the near-space mesh network with the massive deployment of cost-effective balloons suffers from their undesirable movement caused by winds in the near space, as illustrated in Fig. \ref{fig4}.} Since there possibly exist balloon nodes ``pushed'' away from its working region by wind effects, the placement of near-space balloons has to be adjusted correspondingly to maintain seamless access of ground users, despite their limited capability of flight control. To this end, Google cleverly utilized the winds at different altitudes with various strength and directions, to control the aviation of near-space balloons. Explicitly, based on wind forecast information, the altitude of balloons were adjusted through inflation and deflation, to capture the winds with desirable speed and directions. Then the balloons can be sent to the wanted destination. Aside from ensuring coverage of the ground users, the dynamic networking and routing algorithms are also essential to maintain the connectivity between different balloon nodes with low mobility. {For instance, the deployment of redundant NSPS nodes and inter-NSPS data paths together with dynamic link rewiring strategy can enable resilience against possible paralysis of the balloon nodes, where the deploying cost issue also requires careful consideration.}

{ Note that the near-space mesh network can be also established by hybrid airship and balloon constellations. To guarantee its performance, such as coverage, data throughput and reliability, based on the discussions above, we need to jointly optimize the placement topology of static airship nodes, and the dynamic networking/routing strategies against mobility of balloon nodes, where the hierarchical level of the airship nodes is set to be higher than the balloons.}

{\bf Co-existence with Terrestrial Networks}: The extended coverage of NS-COM could induce non-negligible interference with ground cellular networks, leading to performance loss of both the near-space and ground networks. To tackle this issue, we may dynamically manage the frequency resources allocated for the cells of NS-COM and terrestrial networks, which requires advanced spectrum sensing algorithms. However, such a spectrum management scheme may be unpractical under the scenario of crowded data traffic. Another feasible strategy is adjusting the beam directions of NSPS to point away from the terrestrial cell, in order to mitigate the interference, which can be realized by the smart antenna technologies. Besides, the well-known inter-cell interference coordination (ICIC) and coordinated multi-point (CoMP) schemes considered in LTE/LTE-A can be also invoked for interference elimination under the scenario of spectrum sharing between the terrestrial networks and NS-COM.

Instead of merely treating each other as the interferer, the terrestrial cellular networks and NS-COM can be integrated to inherit the advantages of both tiers, and boosting their respective performance by efficient coordination. The near-space-ground integrated network is constituted by the near-space mesh network and the ground network, which are connected with the near-space access network. {This complex networking architecture necessitates joint design spanning from physical layer to link layer and network layer, in order to ensure the overall performance of the integrated network, including the data throughput, latency, reliability, energy consumption, etc. For instance, regarding the physical layer, the network throughput and reliability can be balanced by cleverly optimizing the center frequency, bandwidth, modulation formats and waveform design for different data links according to their specific propagation environment and QoS requirement.} For the link layer, an adaptive handover strategy between the ground and near-space base stations is necessary for performance enhancement in terms of the latency and reliability, which needs to be designed by considering the weather condition, LoS link availability and the mobility of users. {Besides, traffic offloading with NSPS constellations helps to alleviate crowded bandwidth usage of the terrestrial cellular network, leading to improved data throughput.} Finally, for the network layer, unified design of the routing protocol for the heterogeneous near-space-ground network is critical to avoid under-utilization of the networking resources, which can attain higher energy efficiency.


\section{Candidate Techniques and Challenges}\label{S4} 
In this section, we investigate several candidate communication techniques for NS-COM from the perspective of spectrum usage, followed by a discussion about the potential research opportunities therein.

{\bf Sub-6GHz Communications}: Sub-6 GHz has long been employed for wireless communications with mature communication technologies, which provides continuous coverage and reliable data links. Besides, sub-6GHz signals experience lower path loss, and are less likely to be affected by the propagation environment compared with higher-frequency counterparts, making it suitable for deployment on NSPS operating at the altitude over 20 km. However, there are several drawbacks for the use of sub-6 GHz in NS-COM. Firstly, it is unrealistic to support high-rate services due to its narrower bandwidth. Secondly, the sub-6 GHz band is highly crowded for mobile cellular systems, Wi-Fi, Bluetooth and other wireless communication systems at this frequency band, which may cause severe interference issue. Therefore, the frequency resource management as well as interference mitigation schemes are necessary for implementations of the sub-6GHz-based NS-COM. By considering its features of high signal quality and low data rate, the sub-6 GHz band is suitable for network control signal transmission in NS-COM, scheduling of mmWave links and other low-rate high-reliability services.

{\bf MmWave Communications}: Compared with the sub-6 GHz band, mmWave communications can achieve much higher capacity with sufficient bandwidth up to 252GHz \cite{Zhang_wc_19}. However, the mmWave signals suffer from severe path loss, and are fragile to blockage in the terrestrial environment. Fortunately, these limitations of mmWave communications are no longer dominant in its application to the NS-COM network. Specifically, the high altitude of the near-space platforms over the stratosphere leads to less probability of blockage of the near-space-to-ground LoS link, as well as marginal attenuation during inter-NSPS transmission, due to low density of the atmosphere. In addition, the beam alignment is easier for the NSPS-mounted mmWave system than its terrestrial counterparts, since the NSPS is capable of maintaining low-mobility or nearly geostationary flying status in the near space. These demonstrate the suitability of mmWave communication systems in the NS-COM network. Nevertheless, there are still several open issues remaining to be addressed. Firstly, the existing mmWave systems are designed for small cells with radius of a few hundred meters or less. In contrast, the ground coverage radius of NS-COM systems can reach at least tens of kilometers, which necessitates revolutionary design of mmWave systems in the NS-COM network. Secondly, due to the large spatial scale of the near-space-to-ground link, the mmWave channel characteristics can be altered considerably with the decrease of altitude, mainly due to the varying density and constitution of the atmosphere. Therefore, empirical or analytical studies for the mmWave channel of the near-space-to-ground link are necessary. Additionally, despite the low mobility of NSPS, the beam alignment issue cannot be underestimated due to the narrow beam width of mmWave signals, where advanced beam tracking algorithms are required to enhance the reliability.

{\bf Terahertz Communications}: The exploitation of frequency resources can be further extended to THz frequencies (0.1-10 THz), which provides sufficient available bandwidth to meet the escalating demands on fast data transfer over 100 Gb/s of various bandwidth-consuming services, e.g., holographic video conferencing and wireless backhaul. Short-range THz wireless communications are mainly considered in the terrestrial environments, which suffers from severe molecular absorption by water vapor. On the contrary, the attenuation effects are significantly reduced in the near space with scarce density of water vapor. Besides, the extremely-short wavelength of THz waves enables construction of ultra-massive antenna array, leading to higher beamforming gain than lower frequencies. These make THz communications definitely suitable for inter-NSPS long-distance transmission. Aside from data transmission, the THz signals are also qualified for surveillance missions of NS-COM in the airborne area and above, which can achieve superior resolution of target range and velocity estimation, due to the ultra-high carrier frequency and signal bandwidth. However, it is still quite challenging for practical implementations of THz communications in NS-COM. {Firstly, the inhomogenous atmospheric propagation medium in the near space, as illustrated in Fig. \ref{fig2} leads to undesirable refraction effects especially for ultra-long-range transmission over 100 km, which is detrimental for THz beam alignment due to its ``razor sharp'' beam shape.} Secondly, the use of THz-band large-scale antenna array induces sophisticated delay and beam squint effects \cite{Liao_jsac_21}, leading to increased inter-symbol interference (ISI) and undesirable beam split effects. {Besides, the extremely high center frequency over 100 GHz leads to more severe Doppler shift than that at mmWave frequencies, which presents frequency-selective property due to the ultra-broad communication bandwidth, yielding the so-called Doppler squint effect.} In addition, the hardware imperfection issue, usually neglected at low frequencies, cannot be overlooked for NSPS-mounted THz transceiver devices, which imposes nonlinear distortions on the THz signals. This is because there is much difficulty for hardware fabrication in the well-known ``THz gap'' between microwave and infrared frequencies. Besides, NSPS tends to utilize direct-conversion architecture for its low complexity and expenses, which is more sensitive to hardware imperfections. The resultant distortions cannot be addressed with ease using the existing compensation/calibration methods, due to their sophisticated wideband feature and the limited payload budget for communication equipment on NSPS.

{\bf Free-Space Optical Communications}: The free-space optical (FSO) communications offers sufficient unlicensed bandwidth for high rate transmission, which is fragile to atmospheric turbulence and absorption effects. These factors are marginal in the near space with scarce density of atmosphere, making FSO communications a promising candidate for the inter-NSPS links in NS-COM. Furthermore, the beam divergence issue for inter-satellite communications caused by small beam width \cite{Chaudhry_cem_21} is alleviated under inter-NSPS propagation scenario, thanks to the low-mobility nature of the aerostatic near-space platforms. Nevertheless, the application of FSO communications inevitably suffers from performance degradation caused by limited transmit power under sky radiance and background shot noise, which requires advanced transceiver design to maintain the system error performance. Note that the mathematical signal model may be completely different from the classical RF counterparts due to the introduction of Poisson-distributed shot noise components.

\section{Conclusion}\label{S5}
In this article, we provided an extensive overview of near space communications (NS-COM) by specifying its differences in comparison with the other space/air/ground networks, and introducing its research development. Then the peculiarities of NS-COM were summarized from two aspects of the NSPS nodes and the near-space propagation environment, and the corresponding technical issues were highlighted, which were mainly raised by the ultra-long transmission distance, limited volume/weigh for communication payloads, and {non-homogeneous} atmospheric environment of the near space. Based on the characteristics of NS-COM, we discussed several application scenarios of NS-COM, i.e., the near-space-to-ground P2P link, the near-space access network and the near-space mesh network, and revealed their technical requirements ranging from the physical-layer aspect of modulation and beamforming to the networking aspect of data offloading and multi-NSPS deployment. Furthermore, the co-existence of NS-COM and terrestrial cellular networks was investigated in two scenarios, i.e., treating each other as interferers and being integrated together as collaborators. Finally, we also listed several potential techniques for NS-COM in terms of the frequency band, and pointed out the obstacles to their practical implementation.


\begin{thebibliography}{1}
\bibitem{Zhang_conf_07}
Z. Bo, R. Qinghua, L. Yunjiang, C. Zhenyong, and Z. Feng, ``Characteristic and simulation of the near space communication channel,'' {in \emph{Proc. MAPE 2007}}, Hangzhou, China, Aug. 2007, pp. 769-773. 

\bibitem{Abbas_procIEEE_11}
A. Mohammed, A. Mehmood, F. Pavlidou, and M. Mohorcic, ``The {role} of high-altitude platforms (HAPs) in the global wireless connectivity,'' \emph{Proc. IEEE}, vol. 99, no. 11, pp. 1939-1953, Nov. 2011.

\bibitem{Chan_cm_16} 
S. Chandrasekharan, K. Gomez, A. Al-Hourani, S. Kandeepan, T. Rasheed, L. Goratti, L. Reynaud, D. Grace, I. Bucaille, T. Wirth, and S. Allsopp, ``Designing and implementing future aerial communication networks,'' \emph{IEEE Commun. Mag.}, vol. 54, no. 5, pp. 26-34, May 2016.

\bibitem{Alam_cm_21}
M. S. Alam, G. K. Kurt, H. Yanikomeroglu, P. Zhu, and N. D. Dao, ``High altitude platform station {based} super macro base station constellations,'' \emph{IEEE Commun. Mag.}, vol. 59, no. 1, pp. 103-109, Jan. 2021.

\bibitem{StationX_2022} 
\emph{X-Station}. Accessed: Mar. 10, 2022. [Online]. Available: \url{http://www.stratxx.com/xstation.html}

\bibitem{Loon_Google_2022} 
\emph{Loon: Expanding internet connectivity with stratospheric balloons}. Accessed: Mar. 10, 2022. [Online]. Available: \url{https://x.company/projects/loon/}

\bibitem{Liu_survey_18} 
J. Liu, Y. Shi, Z. M. Fadlullah, and N. Kato, ``Space-air-ground integrated network: A survey,'' \emph{IEEE Commun. Surveys \& Tut.}, vol. 20, no. 4, pp. 2714-2741, Fourthquarter 2018.


\bibitem{Ozdemir_conf_13}
M. C. Ozdemir, ``Conceptual changes by use of near space,'' in \emph{Proc. IEEE/AIAA 32nd Digital Avionics Systems Conference (DASC)}, East Syracuse, NY, USA, Oct. 2013, pp. 3E3-1-3E3-10.

\bibitem{britannica_2022} 
\emph{Layers of Earth's atmosphere}. Accessed: Mar. 24, 2022. [Online]. Available: \url{https://www.britannica.com/science/atmosphere/Troposphere#/media/1/41364/99826}

\bibitem{Pang_wc_21}
{X. Pang, M. Sheng, N. Zhao, J. Tang, D. Niyato, and K. -K. Wong, ``When UAV meets IRS: Expanding air-ground networks via passive reflection,'' \emph{IEEE Wirel. Commun.}, vol. 28, no. 5, pp. 164-170, Oct. 2021.}

\bibitem{Tang_wc_20}
W. Tang, M. Chen, J. Dai, Y. Zeng, X. Zhao, S. Jin, Q. Cheng, and T. Cui, ``Wireless communications with programmable metasurface: New paradigms, opportunities, and challenges on transceiver design,'' \emph{IEEE Wirel. Commun.}, vol. 27, no. 2, pp. 180–187, Apr. 2020.

\bibitem{Gong_tec_17} 
M. Gong, Z. Wang, Z. Zhu, and L. Jiao, ``A similarity-based multiobjective evolutionary algorithm for deployment optimization of near space communication system,'' \emph{IEEE Trans. Evolutionary Computation}, vol. 21, no. 6, pp. 878-897, Dec. 2017.	

\bibitem{Zhang_wc_19}
C. Zhang, W. Zhang, W. Wang, L. Yang, and W. Zhang, ``Research challenges and opportunities of UAV millimeter-wave communications,'' \emph{IEEE Wirel. Commun.}, vol. 26, no. 1, pp. 58-62, Feb. 2019.

\bibitem{Liao_jsac_21}
A. Liao, Z. Gao, D. Wang, H. Wang, H. Yin, D. W. K. Ng, and M. Alouini, ``Terahertz ultra-massive MIMO-based aeronautical communications in space-air-ground integrated networks,'' \emph{IEEE J. Sel. Areas Commun.}, vol. 39, no. 6, pp. 1741-1767, Jun. 2021.

\bibitem{Chaudhry_cem_21}
A. U. Chaudhry and H. Yanikomeroglu, ``Free space optics for next-generation satellite networks,'' \emph{IEEE Consumer Electron. Mag.}, vol. 10, no. 6, pp. 21-31, Nov. 2021.







	
\end{thebibliography}
\end{document}